\documentstyle[12pt] {article}
\def\beq{\begin{equation}}
\def\eeq{\end{equation}}
\def\bea{\begin{eqnarray}}
\def\eea{\end{eqnarray}}
\def\bq{\begin{quote}}
\def\eq{\end{quote}}
 
\begin{document}
\pagestyle{empty}
\begin{flushright}
\end{flushright}
\vspace*{5mm}
\begin{center}
{\bf EXACT BETA FUNCTION FROM THE HOLOGRAPHIC LOOP EQUATION
OF LARGE-$N$ $QCD_4$}
\\  
\vspace*{1cm} 
{\bf M. Bochicchio} \\
\vspace*{0.5cm}
INFN Sezione di Roma \\
Dipartimento di Fisica, Universita' di Roma `La Sapienza' \\
Piazzale Aldo Moro 2 , 00185 Roma  \\ 
e-mail: marco.bochicchio@roma1.infn.it \\
\vspace*{2cm}  
{\bf ABSTRACT  } \\
\end{center}
\vspace*{5mm}
\noindent
 We construct and study a previously defined quantum holographic
 effective action, $\Gamma_q$, whose critical equation implies the
 holographic loop equation of large-$N$ $QCD_4$ for planar self-avoiding
 loops in a certain regularization scheme.
 We extract from $\Gamma_q$ the exact beta function in the given scheme.
 For the Wilsonean coupling constant the beta function is exactly one loop
 and the first coefficient, $\beta_0$, agrees with its value in perturbation
 theory.
 For the canonical coupling constant the exact beta function has a $NSVZ$ form
 and the first two coefficients in powers of the coupling,
 $\beta_0$ and $\beta_1$, agree with their value in perturbation theory.
\vspace*{1cm}
\begin{flushleft}
\end{flushleft}
\phantom{ }
\vfill
\eject
%\pagestyle{empty}
%\clearpage\mbox{}\clearpage
\setcounter{page}{1}
\pagestyle{plain}

\section{Introduction}

From a purely computational point of view, one of the aims of this
paper is to show that the exact beta function of large-$N$
$QCD_4$, for the 't Hooft canonical coupling $g^2=g^2_{YM} N$ that occurs
in our effective action, is given by:
\bea
\frac{\partial g}{\partial log \Lambda}=\frac{-\beta_0 g^3+
\frac {\beta_J}{4} g^3 \frac{\partial log Z}{\partial log \Lambda} }{1- \beta_J g^2 }
\eea
in a certain regularization scheme to be specified later in this paper,
with:
\bea
\beta_0=\frac{1}{(4\pi)^2} \frac{11}{3} \nonumber \\
\beta_J=\frac{4}{(4\pi)^2} 
\eea
and $Z$ to be defined below.
At the same time, the beta function for the 't Hooft Wilsonean coupling,
that occurs in our effective action, is exactly one loop:
\bea
\frac{\partial g_W}{\partial log \Lambda}=-\beta_0 g_{W}^3
\eea
While the mentioned scheme is not easy to compare 
to anyone which may be chosen in perturbation theory,
perhaps the most relevant aspect of
Eq.(1) is that it is deduced from a version of the large-$N$
loop equation for planar self-avoiding loops and that, once
the lowest order result for $Z$:
\bea
Z=1+g^2 \frac{1}{(4\pi)^2} \frac{10}{3} log(\frac{\Lambda}{\mu})
\eea
is inserted in Eq.(1), it implies the correct value of the first and
second perturbative coefficients of the beta function \cite{AF,AF1,2loop1,2loop2}:
\bea
\frac{\partial g}{\partial log \Lambda}=
-\beta_0 g^3+
(\frac {\beta_J}{4} \frac{1}{(4\pi)^2} \frac{10}{3} -\beta_0 \beta_J) g^5 +... \nonumber \\
=-\frac{1}{(4\pi)^2}\frac{11}{3} g^3 + \frac{1}{(4\pi)^4} ( \frac{10}{3}
-\frac{44}{3})g^5 +... \nonumber \\
=-\frac{1}{(4 \pi)^2} \frac{11}{3} g^3 -\frac{1}{(4 \pi)^4} \frac{34}{3} g^5+...
\eea
which are known to be universal, i.e. scheme independent.
In a different large-$N$ limit of $QCD_4$, a different exact beta function of $NSVZ$ type
was obtained in \cite{A1,A2}, by finding a link with the large-$N$
$ \cal{N}$ $=1$ $SUSY$ gauge theory. It was then argued \cite{A1,A2} that also the ordinary
large-$N$ 't Hooft limit considered in this paper admits $SUSY$ relics
and in particular a beta function of $NSVZ$ type. It is possible that the preceding observation
explains why the exact beta function found in this paper has a $NSVZ$ structure,
despite the absence of any super-symmetry. In this respect, but more intrinsically from the point
of view of this paper, it might be relevant the existence, in the pure large-$N$ $YM$ theory,
of an analogue of the chiral
ring of the $ \cal{N}$ $=1$ $SUSY$ gauge theory, mentioned later in this introduction.
The exact beta function of Eq.(1) follows from our construction of the holographic effective
action, which is in fact the aim of this paper from a broader
point of view.
Before showing the details, we should perhaps
mention in which sense our construction solves the loop equation in the
large-$N$ limit.
Usually, by the solution of the large-$N$ limit it is meant finding
an operator valued connection $A_{\alpha}$, the master field \cite{W}, living in some
non-commutative type $II_1$ von Neumann algebra (i.e. the algebra has a finite normalised trace) \cite{Si},
that solves the following loop equation \cite{MM,MM1} uniformly for all loops:
\bea
0=\int DA_{\alpha}\exp(-\frac{N}{2 g^2} \sum_{\alpha \ne \beta} \int Tr( F_{\alpha \beta}^2) d^4x)
(Tr(\frac{N}{ g^2} D_{\alpha} F_{\alpha \beta}(z) \Psi(x,x;A)) \nonumber \\
 +i\int_{C(x,x)} dy_{\beta} \delta^{(d)}(z-y) (Tr( \Psi(x,y;A)) Tr(\Psi(y,x;A)))
\eea 
with:
\bea
\Psi(x,y;A)=P \exp i\int_{C_{(x,y)}} A_{\alpha} dx_{\alpha}
\eea
This is a very difficult problem, since the ambient algebra of based Wilson loops in 
the large-$N$ limit
is a non-hyperfinite von Neumann algebra, i.e. an algebra that is not
the limit of a sequence of finite dimensional matrix algebras \cite{MB1,G,D,D1,V}.
By no means our construction solves this problem, that is essentially equivalent to find
the exact $1PI$ effective action in the large-$N$ limit \cite{Haa1,Haa2,Cv}.
Rather, we consider solving the loop equation for a fixed planar self-avoiding loop.
Since the loop is fixed, the corresponding algebra, obtained iterating the loop,
is commutative \cite{MB1,Duh}. In addition the connection, whose holonomy is computed by our choice of the
Wilson loop, is of a special type. Its curvature is a linear combination of the anti-selfdual (ASD)
components only. In a sense that will become apparent in the rest of the paper, our Wilson loop
belongs to an $ \cal{N}$ $ = 0 $ analogue of the chiral ring \cite{Vafa,CD} 
of $ \cal{N}$ $ = 1$ super-symmetric gauge theories \cite{Ve}.
Thus we are looking for the solution of a much simpler, yet non-trivial
problem, that contains a more limited, but still very interesting information.
The version of the large-$N$ loop equation that we refer to,
has been named holographic by us in \cite{MB1}, because it involves
a boundary-bulk correspondence between the loop equation of
large-$N$ $QCD_4$, which
lives on loops, and a holographic effective action
whose critical equation, which lives on points, implies the loop equation
in its holographic form.
We should justify better why we named holographic such construction.
The loop equation can be roughly seen as the sum
of a classical term, that is the easy one to control, because it has already
the form of a critical equation for an effective action  
(the classical one indeed) and a quantum term, the difficult one to control,
because it is a contour integral
along the loop and thus gives the loop equation a structure very different
from a critical equation defined on points.
Now, loosely speaking, there is a way to associate to a loop a point,
via the evaluation of a residue.
In fact the Cauchy theorem can be regarded as the oldest and most remarkable
case of holography.
We implement this idea as follows.
Our strategy, to construct the quantum effective action equivalent to
the loop equation, is to change variables and to make transformations in the loop
equation
in such a way that, in the new variables, the quantum term vanishes
for planar self-avoiding loops whose holonomy is of the special type
mentioned before.
This is achieved in two steps.
In the first step, we change variables in the loop equation from the connection to
its curvature, in such
a way that the quantum term, that is a contour integral, is reduced
to the computation of a regularized residue, evaluated at any marked point
of the loop that enters the loop equation.
The reason for which this is possible for our special choice of the connection 
is that a holomorphic gauge exists, in which functionally differentiating
the connection with respect to its curvature in the loop equation
produces the Cauchy kernel.
In the second step, the region inside and the one outside the marked loop 
are mapped by a conformal transformation to two cuspidal fundamental domains
(we get control over the cusp anomaly) in the upper half plane, in such a way
that to any marked point of the loop are attached infinitesimal strips
ending into the cusps at infinity. Attaching to the marked points the
infinitesimal strips does not change the Wilson loop, i.e. the holonomy
of the loop, because of the zig-zag symmetry. The zig-zag symmetry
means that the Wilson loop is left invariant if any arc that backtracks
is added to the loop \cite{Pol2}.
The regularized residue vanishes at the cusps (i.e the image
of the marked points by the conformal map)
because of the zig-zag symmetry of the loop in a neighbourhood of the cusps,
thus implying the existence
of an equivalent effective action on the conformally transformed domain.
The two domains thus obtained have the loop in common.
We can look at this picture as a hologram of the universe.
This hologram of the universe is in fact enriched by many other
cusps, that can all be chosen to lie on the boundary of the upper
half plane. These cusps are the images in the hologram of a lattice
of points in the (conformally compactified) plane over which the loop lies,
that carry the local degrees of freedom of the gauge theory.
Also these cusps are the end points of infinitesimal strips, starting from the
loop, that can be freely added to the loop without changing the
loop equation because of the zig-zag symmetry.
Thus, in our hologram, all the bulk degrees of freedom of the original theory
live on the boundary line in the upper half plane. 
It has been shown in \cite{MB1} that, from a purely gauge theoretic point of
view, the reason for which
the quantum term in the loop equation vanishes on the hologram
is that on the hologram the structure group of the gauge
theory can be reduced by means of a peculiar gauge fixing:
we can, at the same time, choose an axial gauge in a direction orthogonal
to the line of the cusps and, using the residual gauge symmetry
extended across the cusp line, diagonalize the degrees of freedom that live at the cusps,
to get a theory of $N$ eigenvalues as opposed to the original order of $N^2$
matrix elements. This theory of $N$ eigenvalues is necessarily classical
in the large-$N$ limit and it is determined by the critical equation
of the effective action, thus completing our, by now holographic,
boundary-bulk correspondence.
Therefore holography, for us, is a tool to perform large-$N$ functional integrals.
From this purely gauge theoretic point of view, a subtle point arises about
compactifying the cusps on the conformally transformed domain.
This compactification is absolutely needed:
the loop equation would reduce simply to the classical contribution,
were the marked points not to belong to the loop. 
Because of the compactification, the gauge symmetry must extend to the cusps.
However, extending the gauge freedom to the cusps creates in general a Dirac string
and it is not compatible with the geometry of the cusps as parabolic points.
In fact the arcs ending into the cusps associated to the usual parabolic points
share the same orientation in order to form tubes, after pairwise identification,
ending into the cusps at infinity.
The Dirac string forbids the identification 
of the two arcs and thus the extension of the gauge symmetry to the cusps,
as it should be for truly parabolic points. Indeed the moduli of parabolic bundles
are usually defined requiring that the gauge group acts trivially at the cusps \cite{Witt}.
Yet, if the arcs ending into the cusps have opposite orientation, as implied by the loop orientation,
the existence of the Dirac string is compatible with the opposite orientation of the arcs,
because these arcs must not be identified: indeed they form strips and not tubes.
Thus, in this case, the gauge symmetry can be extended to the cusps.
In this paper we choose a Wilson loop in the adjoint representation, that in the large-$N$
limit factorizes into the product of two Wilson loops in the fundamental representation and 
its conjugate.
Since the v.e.v. of a Wilson loop does not depend on its orientation and on which between 
the fundamental representation or its conjugate is chosen,
at global level there are two possibilities of gluing the
two charts of the hologram along the boundary loop.
They correspond to an orbifold or to an orientifold \cite{Orb,Ve2}.
In $d=4$, we present our construction for the orientifold case only \cite{Ve2}.
The essential reason is that any marked point of the loop has in fact two images, one for each
of the two hologram charts. Therefore the infinitesimal strips added to the loop must occur in pairs,
one inside and one outside the loop. Thus also the cusps occur in pairs,
so that the lattice inside and the one outside the loop have the same number of cusps.
Hence the theory needs necessarily two different lattice scales, $\tilde a$ and $a$, that are used
to measure, for example, different areas, since the number of lattice points in the two charts
is the same.
Thus space-time does not contain a lattice of uniform spacing.
This is somehow irrelevant in the $d=2$ theory, but introduces great computational
difficulties in $d=4$. These difficulties persist in the orbifold construction.
The orientifold construction, instead, merges hologram charts with 
the same lattice spacing, but with conjugate representations of the connection,
allowing explicit computations.
Let us describe in more detail the holographic loop equation \cite{MB1}.
The loop equation in its conventional form is written in terms
of a generic non-planar Wilson loop. However to be able to evaluate
the quantum term as a residue we need a planar Wilson loop.
Planar unitary holonomies are not a complete system of observables
in the $d=4$ theory. To partially fix this, we consider planar non-unitary Wilson loops,
built by means of a non-hermitean connection whose curvature
is a certain linear combination of the $ASD$ part of the curvature only.
Correspondingly, to get our new form of the loop equation,
we introduce a resolution of identity in the functional integral
into the levels of the $ASD$ part of the curvature.
To be compatible with this resolution, we employ the well known decomposition of
the classical action into a topological term, the second Chern class,
and a term containing only the $ASD$ part of the curvature.
Our loop equation is written using as integration variable the $ASD$ 
part of the curvature. The second Chern class does not contribute to the loop
equation. Yet, to get a planar theory in space-time and thus a residue in the loop equation,
we need a non-commutative (in a plane orthogonal to the plane of the Wilson loop)
Eguchi-Kawai ($EK$) reduction of the theory from four to two dimensions,
in the limit of infinite non-commutativity, that is equivalent to the original theory
in the large-$N$ limit \cite{EK,Neu,Twc,Twl1,Twl2}. This reduces the $ASD$ part of the curvature 
to a curvature of Hitchin type \cite{Hit}, but some of the global four dimensional
information survives in the second Chern class and in the existence of a central extension in the $ASD$
part of the curvature due to non-commutativity, that is related to the first Chern class.
We then require the local part of the curvature of Hitchin type to be localised into a linear combination
of two dimensional delta functions. This realizes many purposes.
It gives us a dense basis valued in the distributions, for integrating
over the curvature in the functional integral. It gives us a nice moduli space
\cite{S1,S2,S3,S4,Biq}.
It gives us a curvature localised on points, to which we can attach infinitesimal
strips in the loop equation. It gives us nice formulae for the first and second
Chern class in terms of parabolic Higgs bundles \cite{S4,Witt}. These
Chern classes depend also on four dimensional features
of the fibration of the parabolic Higgs bundles, for example intersection numbers.
This four dimensional information is given independently of the loop
equation and it is represented by a discrete set of choices some
of which must be compatible with the global structure of the hologram. \\
The content of this paper is as follows. 
In sect.2 we explicitly construct
the quantum holographic effective action, $\Gamma_q$, an object that was
previously defined in \cite{MB1} through an auxiliary
quantity, $\Gamma$, the classical holographic effective action.
In fact for technical reasons, to construct the large-$N$ limit, we employ a twisted version of the pure
$YM$ theory,
that is equivalent to a non-commutative theory on  $R^2 \times {R^2}_{\theta}$ in the limit of infinite
non-commutativity $ \theta \rightarrow \infty $.
Sect.2 contains several refinements
with respect to \cite{MB1}. In particular it employs a well known
decomposition of the classical $YM$ action into the second Chern class
and a purely $ASD$ term, that is particularly well suited both theoretically
and computationally for
getting our holographic effective action, but that was not considered
in \cite{MB1}.
We give also a uniform
treatment of the cusps
corresponding to the marked points of the loop and the ones corresponding
to the remaining points, an improvement with respect to \cite{MB1}.
We also mention the absence of a logarithmic cusp anomaly
in the loop equation in $d=4$.
Finally, we point out some unexpected links with the operad structure of the arc complex
defined over Riemann surfaces with boundaries \cite{Pen}.
In sect.3 we write down the holographic effective action in $d=2$
$(d=4)$, taking into account in our hologram both regions delimited by the Wilson
loop on a (large) sphere
and the related fact that the eigenvalues of the curvature of the (twisted) local
system that enters the construction are defined only up to some shifts.
The occurrence of these shifts in the curvature of the connection
is explained by the fact that the logarithm of the eigenvalues of the holonomy of the connection
around the loop are determined only modulo $2\pi i$ and that 
the two charts of the hologram are glued together summing over field configurations that
keep the holonomy of the connection around the loop fixed.
These shifts are important both in two and four dimensions. In two dimensions
they lead to the confining strong coupling phase transition \cite{GW,D2,KA}.
In four dimension presumably they lead to confinement as well.
The occurrence of the shifts may be a hint of the existence
of a stringy representation of the partition function in the $d=4$
case at least for large loops, following the analogy with the $d=2$ theory in the strong
coupling phase \cite{D2,KA}, where on the string side they can be related
to the winding of the string around the loop \cite{D2,KA}.
Since we compute the effective action for a Wilson loop in the adjoint representation,
we are looking effectively to a $SU(N)/Z_N$ theory.
Therefore the partition function contains a sum with equal weights over sectors of $Z_N$ flux.
This implies that our quantum holographic effective action contains also a sum over these sectors as well.
In sect.3 we compute exactly the local part of the holographic effective action (in the language of
$AdS$ holography
this is the near horizon limit \cite{Mal}), up to finite terms and up to a conformal anomaly. 
An essential feature of the orientifold case for the adjoint representation
is that our computation almost factorizes into the product of two contributions
associated to two different holograms.
One hologram is the orientifold obtained merging the regions close to the point at infinity.
We call it the hologram at infinity.
The other one is obtained merging the regions close to zero, the antipodal point on a sphere.
We call it the hologram at zero.
In sect.3 we compute the beta function of the Wilsonean coupling constant,
finding that it is exactly one loop in the given scheme and that agrees with the one-loop
perturbative result.
We compute also the exact beta function for the canonical coupling,
finding exact agreement with the one and two loop perturbative result, that is known to be
scheme independent.
The exact canonical beta function, as opposed to the two loop perturbative one, has,
in addition to the usual perturbative ultraviolet fixed point, an infrared fixed point.
Indeed the infrared fixed point occurs at the value of the coupling
for which the numerator in Eq.(1) has a zero, as a consequence of the
cancellation between terms of different orders in $g$ with opposite
signs. The value of the running coupling for which the infrared zero of the beta function
occurs is scheme dependent, due to
the scheme dependence of higher order coefficients of $ \frac{\partial log Z}{\partial log \Lambda} $. 
In sect.4 we recall our conclusions.

\section{Holography as a tool to perform large-$N$ functional integrals:
the quantum holographic effective action}

Our starting point is the pure $SU(N)$ $YM$ theory defined over the four manifold
$R^4$. Our observable is a Wilson loop in the adjoint representation in Euclidean 
space-time. We will eventually perform the analytic continuation of the Euclidean 
Wilson loop to Minkowskian space-time and also to ultra-hyperbolic signature.
In this case we obtain a Wilson loop on the light cone in Minkowskian signature
or on a diagonally embedded light cone in ultra-hyperbolic signature.
The partition function reads:
\bea
Z=\int \exp(-\frac{1}{4g^2} \sum_{\alpha \neq \beta} \int Tr_{adj}(F_{\alpha \beta}^2) d^4x) DA
\eea
where the generators of the Lie algebra in the adjoint representation are normalised as:
\bea
Tr(T^a T^b)_{adj}= N \delta_{ab} \nonumber \\
\sum _a (T^a)_{adj}^2=N 1_{adj}
\eea
In the large-$N$ limit it is convenient to factorize the v.e.v of a Wilson loop in the adjoint
representation into the product of the v.e.v. in the fundamental representation and its complex conjugate.
The corresponding partition function factorizes into:
\bea
Z=\int \exp(-\frac{N}{2g^2} \sum_{\alpha \neq \beta} \int Tr_f(F_{\alpha \beta}^2) d^4x) DA  \nonumber \\
\times
\int \exp(-\frac{N}{2g^2} \sum_{\alpha \neq \beta} \int Tr_{\bar f}(\bar{F}_{\alpha \beta}^2) d^4x) D\bar{A}
\eea
where the generators in the fundamental representation and its conjugate are normalised as:
\bea
Tr_{f}(T^a T^b)= \frac{1}{2} \delta_{ab} \nonumber \\
\sum_a (T^a)_{f}^2=\frac{N^2-1}{2 N} 1_{f}
\eea
For notational convenience and brevity, we perform our analysis of the loop equation for just
one factor, for example the one corresponding to the fundamental representation.
This suffices to write down the holographic form of the loop equation on the original space-time
and to show the vanishing of the quantum contribution in the loop equation on each chart of
the corresponding hologram at local level. However, at global level, to perform our orientifold
construction, we are required to put together both the fundamental and anti-fundamental factors 
of the adjoint representation.
The large-$N$ limit of the pure $YM$ can be reduced to a planar problem by a partial
$EK$ reduction, in which two, among the four space-time dimensions, are
reduced to a point. Thus our large-$N$ theory, in its continuum version,
is a non-commutative theory on $R^{2} \times R^{2}_{\theta}$ in the limit
of infinite non-commutativity. This is the twisted theory.
Then the degrees of freedom corresponding to the non-commutative $R^2_{\theta}$ are absorbed into the colour Hilbert space.
This is the twisted reduced theory, that is two dimensional, but with fields
living into an infinite dimensional colour space. This means, for example,
that the derivatives in the non-commutative directions are interpreted as creation and annihilation operators
in an infinite dimensional matrix representation in colour space.
It is well known, directly from the loop equation or by functional integral methods \cite{RT},
that the classical action of the theory thus reduced must be rescaled by a factor of
$N_2^{-1}$, to compensate the reduction of the entropy in the functional integration.
$N_2$ represents the number of semi-classical quantum states in the directions 
transverse to the loop.
In the ordinary theory on commutative space-time $N_2=\frac{1}{(2 \pi)^2}\int d^2x d^2p$.
In the non-commutative case $N_2$ satisfies $ \frac{2 \pi}{N_2 H \tilde a^2}=1$, where $H$ is the inverse
of the non-commutative parameter, $\theta$, and $\tilde a$ the lattice cutoff.
Let us notice that when $H$ is normalised 
as $H=\frac{2\pi}{V_2}$, with $V_2$ the area of the transverse space-time, $N_2= \frac{V_2} { \tilde a^2 }$
as in the commutative case.
The reduction process could be continued until the four dimensional
theory is reduced to a $0$-dimensional matrix model,
but, for our purposes, we require only a partial $EK$ reduction to two dimensions.
We now give a heuristic description of how the holographic map works
for the reduced theory.
We are given a two dimensional gauge theory with a residual two dimensional
gauge symmetry.
In a lattice version this theory is a reduced twisted $EK$
model, in which the local gauge degrees of freedom live on the links
of a planar lattice. From the point of view of the large-$N$ functional
integration each link carries order of $N^2$ integration variables
and thus the functional integral cannot be performed by the saddle point method,
because the entropy is of the same order of $N^2$.
Let us suppose that, in some way, we can pass from the links of the planar lattice to the
points of the dual lattice.
In the language of the continuum theory we pass from the connection to
the curvature as fundamental integration variable. 
In the $d=2$ theory, this is simply the curvature of the connection that enters the 
functional integral.
In the $d=4$ theory, this is actually the $ASD$ part of the curvature in a version of the
twisted reduced $EK$ model.
The corresponding observable is a Wilson loop with a non-hermitean holonomy, whose
curvature coincides with a non-hermitean combination of the $ASD$ part of
the curvature in the Euclidean signature. 
However, the analytic continuation to ultra-hyperbolic signature can be performed 
in such a way that the curvature becomes hermitean.
The ultra-hyperbolic signature is obtained by analytic continuation
from Euclidean to Minkowskian space-time 
taking into account that the gauge invariant regularization of the loop equation that
we will employ requires analytic continuation from Euclidean to
Minkowskian space-time and, as result, the planar Wilson loop analytically continued
lives on a light cone, diagonally embedded in $ R^2 \times R_{\theta}^2 $.
The change of variable from the connection to the $ASD$ part of the curvature
defines the classical holographic
effective action, $ \Gamma  $, that is obtained by adding to the classical action the logarithm
of the Jacobian of the change of variable
from the connection to the $ASD$ curvature, plus the logarithm of another Jacobian,
due to the choice of a holomorphic gauge chosen in order to get the Cauchy kernel
in the loop equation.
In $d=4$, the $ASD$ part of the curvature suffices to resolve the identity in the functional integral,
since the classical action is written as the sum of the $ASD$ part and of a 
topological term involving the (parabolic) second Chern class, that is kept fixed by
quantum fluctuations, and that, for the stable parabolic Higgs bundles introduced momentarily, vanishes
identically \cite{S4}.
Yet, at each point of the dual lattice, we have still order of $N^2$ integration
variables. Hence the name classical holographic action for $\Gamma$,
since $\Gamma$ defines the classical action still to be integrated
in the functional integral over the $ASD$ curvature.
Now we compactify the $d=2$ space time of the reduced theory
to a sphere,
requiring that all the fields approach a definite limit at
infinity.
Our dual lattice defines a divisor on the sphere,
on which the curvature of our twisted infinite dimensional parabolic 
Higgs bundle is localised. 
The twist refers here to a constant central term occurring in the $ASD$ part
of the curvature, in addition to the delta-like singularities,
and is due to the non-commutative nature of the twisted $EK$ reduction
in the continuum limit.
By the uniformization theory, the sphere with punctures is conformally equivalent
to a cuspidal fundamental domain in the upper half plane, whose parabolic
points are the cusps. This domain of the upper half plane would be, at first sight,
the candidate hologram of the original theory. 
Indeed, on the upper half plane, that is the universal cover, we can fix an axial gauge
that leaves a residual gauge symmetry along the boundary,
that is the line where the cusps sit. Could we extend the gauge symmetry to the
cusps, using the residual symmetry,
we could impose an extra gauge fixing condition in order to reduce the number of integration variables.
However, this does not work for the following reason.
There is a subtle point, that really depends as to whether the Dirac string,
created extending the gauge symmetry across the cusps, is  compatible with the
geometry of the cusps. If the arcs ending into the cusps have the same
orientation, they can be glued to form tubes going to infinity. These
are the cusps that coincide with the usual parabolic points. But then the existence
of the Dirac string is incompatible with gluing, a situation that we could have
anticipated, since we started with parabolic points. Indeed moduli of parabolic
bundles are usually defined requiring that the gauge group act trivially
on the curvature at those points \cite{Witt}.
However, if  the cusps were the ending points of arcs with opposite orientation, these arcs 
could not be identified and thus their geometry would be compatible with the existence of a
Dirac string. Hence, we need a situation in which the arcs ending into the
cusps have opposite orientation.
This can be obtained if strips are added to the marked points of the loop 
in the loop equation. The opposite orientation of the strip sides is then implied
by the loop orientation. But then our hologram has necessarily two charts
with the loop in common.
In this case, the gauge symmetry can be extended to the cusps, and the curvature can
be diagonalized in the hermitean case or triangularized in general (we will see
that, in the loop equation, also a triangular curvature suffices to construct the quantum
effective action; yet, the analytic continuation to Minkowskian space-time, 
that is implicit in the regularization procedure, implies a hermitean curvature).
In the large-$N$ limit this defines a classical theory.
Thus the quantum term in the loop equation has to vanish.
This is the consequence of the zig-zag symmetry along the cusp arcs, that is the same as to say
that the arcs have opposite orientations. \\
Having mentioned the basic ideas, we can now construct in detail our version of the loop equation,
and the corresponding quantum holographic effective action.
Here are the appropriate formulae.
It has been observed sporadically in the literature that the
$YM$ functional integral can be written in the second form 
\bea
Z=\int \exp(-\frac{N}{2g^2} \sum_{\alpha \neq \beta} \int Tr_f (F_{\alpha \beta}^2) d^4x) DA \nonumber \\
=\int \exp(-\frac{N 8 \pi^2 }{g^2} Q-\frac{N}{4g^2} \sum_{\alpha \neq \beta} \int Tr_f(F^{-2}_{\alpha \beta}) d^4x) DA 
\eea
as opposed to the first one.
$Q$ is the second Chern class, given by:
\bea
Q=\frac{1}{16 \pi^2} \sum_{\alpha \neq \beta} \int Tr_f (F_{\alpha \beta} \tilde F_{\alpha \beta}) d^4x 
\eea
with:
\bea
F^-_{\alpha \beta}=F_{\alpha \beta}- \tilde F_{\alpha \beta} \nonumber \\
\tilde F_{\alpha \beta} = \frac{1}{2} \epsilon_{\alpha \beta \gamma \delta} F_{\alpha \beta}
\eea
The last form of the functional integral, though perfectly equivalent
to the usual one, is particularly well suited for the approach to the 
large-$N$ loop equation developed in \cite{MB1} and here.
Indeed, a basic idea in \cite{MB1} is to consider the loop equation
associated to a connection, $B$, whose curvature is of $ASD$ type.
This connection is singled out by the natural choice
of the resolution of identity into the levels $\mu^{-}_{\alpha \beta}$
of the $ASD$ part of the curvature of the gauge connection $A_{\alpha}$:
\bea
1= \int \delta(F^{-}_{\alpha \beta}-\mu^{-}_{\alpha \beta}) D\mu^{-}_{\alpha \beta}
\eea
The partition function thus becomes:
\bea
Z=\int \exp(-\frac{N 8 \pi^2 }{g^2} Q-\frac{N}{4g^2} \sum_{\alpha \neq \beta} \int Tr(\mu^{-2}_{\alpha \beta}) d^4x)
\nonumber \\ 
\times \delta(F^{-}_{\alpha \beta}-\mu^{-}_{\alpha \beta}) D\mu^{-}_{\alpha \beta} DA
\eea
We can write the partition function in the new form:
\bea
Z=\int \exp(-\frac{N 8 \pi^2 }{g^2} Q-\frac{N}{4g^2} \sum_{\alpha \neq \beta} \int Tr(\mu^{-2}_{\alpha \beta}) d^4x)
\nonumber \\
\times Det'^{-\frac{1}{2}}(-\Delta_A \delta_{\alpha \beta} + D_{\alpha} D_{\beta} +i ad_{\mu^-_{\alpha \beta}} ) 
D\mu^{-}_{\alpha \beta} 
\eea
where the integral over the gauge connection of the delta function has been now explicitly performed:
\bea
\int DA_{\alpha} \delta(F^-_{\alpha \beta}- \mu ^-_{\alpha \beta})= |Det'^{-1}(P^- d_A \wedge)|
\nonumber \\ 
=Det'^{-\frac{1}{2}} ((P^- d_A \wedge)^*(P^- d_A \wedge)) \nonumber \\
=Det'^{-\frac{1}{2}}(-\Delta_A \delta_{\alpha \beta} + D_{\alpha} D_{\beta} +i ad_{ F^-_{\alpha \beta}} )
\eea
where $ P^- $ is the projector onto the anti-selfdual part of the curvature and, by an abuse of notation,
the connection $A$ in the determinants denotes the solution of the equation
$F^-_{\alpha \beta}- \mu ^-_{\alpha \beta}=0$.
The $ ' $ suffix requires projecting away from the determinants the zero modes due to gauge invariance, since
gauge fixing is not yet implied, though it may be understood
if we like to. 
We refer to the determinant in Eq.(18) as the localisation determinant because it arises localising the
gauge connection on a given level of the $ASD$ curvature.
Let us notice the somehow unusual spin term $i ad_{ F^-_{\alpha \beta}}$ in Eq.(18).
The non-hermitean connection, $B$, that will enter our loop equation, is somehow adapted to the resolution
of identity:
\bea
B=A+D=(A_z+D_u) dz+(A_{\bar z}+ D_{\bar u})
d \bar z
\eea
$A$ is the projection of the four dimensional hermitean connection
onto the  $(z=x_0+ix_1,\bar z=x_0-ix_1)$ plane of the planar loop
and $D$ is the projection of the four dimensional anti-hermitean non-commutative covariant derivative onto
the orthogonal $(u=x_2+ix_3,\bar u=x_2-ix_3)$ plane.
In this paper we choose the following notation as far as the complex basis
of differentials $dz=dx_0+i dx_1$ and derivatives $\partial=\frac{\partial}{\partial
z}=\frac{1}{2}(\frac{\partial}{\partial x_0}-i\frac{\partial}{\partial x_1})$ is concerned.
Thus, for example, $A_z=\frac{1}{2}(A_0-iA_1)$.
In particular the $ASD$ constraint is interpreted as an equation for the curvature of the
non-Hermitean connection $B=A+D=(A_z+D_u) dz+(A_{\bar z}+ D_{\bar u}) d \bar z$ and a harmonic condition for the Higgs field 
$\Psi=-iD=-i( D_u dz+D_{\bar u} d \bar z )$.
In order to derive our loop equation for $B$, the resolution of identity 
must be rewritten into one of the following formally equivalent forms:
\bea
1=\int \delta(F_B - \mu) \delta(\bar F_{B} - \bar \mu) \delta(d^*_A \Psi - \nu)  D \mu D \bar \mu D \nu 
\eea
or
\bea
1=\int \delta(F_B - \mu) \delta(\bar \partial_A \psi- n)  \delta(\partial_{A} \bar \psi- \bar n) 
D \mu Dn D \bar n
\eea
In the first case, $ D \nu$ is a measure over Hermitean matrices, $\nu=n+ \bar n$,
while $D\mu D \bar \mu$ is a positive measure over complex
matrices. In the second case $Dn D \bar n$ is a positive measure over complex
matrices while $D \mu$ is a complex measure defined as an integral over
the path $\mu=\mu^0+ n-\bar n$ with $\mu^0$ hermitean matrices and $n-\bar n$  kept fixed
while integrating over $\mu^0$. 
The last ingredient, that we need to write down the holographic loop equation,
is the observation that a change of variable exists for the connection
$B$, in which the curvature of $B$ is given by the field $\mu'$,
obtained from the equation:
\bea 
F_B - \mu=0 
\eea
by means of a complexified gauge transformation $G(x;B)$ that puts 
$B=b+ \bar b$ in the holomorphic gauge $\bar b=0$:
\bea 
\bar{\partial}b_z=-i\frac{\mu'}{2}
\eea
where $\mu'=G \mu G^{-1}$.
The mismatch of a factor of $\frac{1}{2}$ between Eq.(22) and Eq.(23) occurs
because Eq.(22) is written in the real basis $dx_0 \wedge dx_1$ while
Eq.(23) is written for the complex components. 
Employing Eq.(21) as a resolution of identity in the functional integral,
the partition function  becomes:
\bea
Z=\int 
 \delta(F_B - \mu) \delta(\bar \partial_A \psi- n)  \delta(\partial_{A} \bar \psi- \bar n) 
 \exp(-\frac{N}{2g^2}S_{YM}) \nonumber \\
 \times \frac{D \mu}{D \mu'} 
Db D \bar b  D \mu' Dn D \bar n
\eea
The integral over $b, \bar b$ is the same as the integral over the four $A_{\alpha}$. The resulting functional
determinants, together with the Jacobian of the change of variables to the
holomorphic gauge, are absorbed into the definition of $\Gamma$.
$\Gamma$ plays here the role of a classical action, since we must
integrate still over the fields $\mu', n, \bar n$.
We may call $\Gamma$ the classical
holographic action, as opposed to the quantum holographic effective action, $\Gamma_q$.
$\Gamma$ for the twisted reduced theory is given by:
\bea
\Gamma=\frac{N 8 \pi^2 }{N_2 g^2} Q
+\frac{N}{g^2} \frac{2\pi}{N_2 H}\int Tr_f(F^{-2}_{01}+F^{-2}_{02}+F^{-2}_{03} ) d^2x \nonumber \\
+ log Det'^{-\frac{1}{2}}(-\Delta_A \delta_{\alpha \beta} + D_{\alpha} D_{\beta} +i ad_{\mu^-_{\alpha \beta}}) 
-log\frac{D \mu}{D \mu'} 
\eea
with:
\bea
\mu^0=F^-_{01}   \nonumber \\
n+\bar n=F^-_{02} \nonumber \\
i(n-\bar n)=F^-_{03}
\eea
The resolution of identity in Eq.(20) is based on a positive measure, while the
one in Eq.(21) is based on a complex measure, that indeed resembles the integration
measure in complex matrix models \cite{Laz}, employed in the study of the chiral ring
of $ \cal {N}$ $ =1 $ $SUSY$ gauge theories \cite{Vafa, CD}. 
The two resolutions of identity are formally equivalent. Yet 
the one in Eq.(21), which has been employed in this paper, contrary to our previous
choice \cite{MB3v1}, leads in natural way to the correct result
for the beta function. Let us explain why. The two choices lead
to different powers of the Vandermonde determinants in the quantum holographic
effective action and also to different ways of counting the dimension of the moduli space of adjoint orbits
and of zero modes.
This is due to a different pairing between the holomorphic and the anti-holomorphic contributions
in the integration measure over the moduli of Higgs bundles. In turn this affects the powers of $g$
that arise by rescaling
the eigenvalues in the Vandermonde determinant and finally it affects the coefficients
of $log g$ in the renormalization of the canonical coupling constant.
Though these differences can be compensated by the different constraints
that arise requiring the vanishing of the parabolic second Chern class, that in turn
affect the normalisation of the classical action, only in the holomorphic case,
as opposed to the hermitean one, we are in fact able to show that there exist some Higgs
bundles for which the parabolic constraints and the counting of zero modes are in fact satisfied
in order to lead to the correct beta function.
By the way, the holomorphic resolution of identity leads to the same powers 
of the Vandermonde determinant that occur in the $d=2$ case.
For completeness we write also the formula for $\Gamma$ in the $d=2$ theory:
\bea
\Gamma=\frac{N}{g^2}\int Tr_f(F^2_{01}) d^2x \nonumber \\
+ log Det'^{-\frac{1}{2}}(-\Delta_A \delta_{\alpha \beta} + D_{\alpha} D_{\beta} +i ad_{\mu_{\alpha \beta}}) 
-log \frac{D \mu}{D \mu'}
\eea
with:
\bea
\mu = F_{01}=\mu_{01} 
\eea
From now on, as far as the loop equation is concerned, we consider only the $d=4$ case,
since the $d=2$ case follows by analogy.
The partition function of the $d=4$ theory is now:
\bea
Z=\int \exp(-\Gamma) D \mu' Dn D \bar n
\eea
In the loop equation it is convenient to consider the Wilson loop as a functional of
the connection $b$, corresponding to gauge transforming $B$ into
the gauge $\bar b=0$. Such a gauge transformation belongs to the
complexification of the gauge group and it is rather a change of variable than
a proper gauge transformation. However, because of the property of the trace,
for closed loops, it preserves the trace of the holonomy. This allows us
to transform the loop equation thus
obtained into an equation for the holonomy of $B$. In our derivation of the loop 
equation, a crucial role is played by the condition that the expectation value
of an open loop vanishes.
In \cite{MB} two slightly different ways of achieving the vanishing of the
expectation value of open $b$ loops were presented. We may thus derive our
loop equation: 
\bea
0=\int  D\mu'  Tr \frac{\delta}{\delta \mu'(w)}
  (\exp(- \Gamma)
  \Psi(x,x;b)) \nonumber \\
 = \int  D\mu' \exp(-\Gamma)
  (Tr(\frac{\delta \Gamma}{\delta \mu'(w)} \Psi(x,x;b)) 
  \nonumber \\ 
  -\int_{C(x,x)} dy_z \frac{1}{2} \bar{\partial}^{-1}(w-y) 
  Tr(\lambda^a \Psi(x,y;b)
  \lambda^a \Psi(y,x;b)) ) \nonumber \\
 =\int D\mu' \exp(-\Gamma)
  (Tr(\frac{\delta \Gamma}{\delta \mu'(w)} \Psi(x,x;b)) \nonumber \\
  - \int_{C(x,x)} dy_z \frac{1}{2} \bar{\partial}^{-1}(w-y)(Tr( \Psi(x,y;b))
    Tr(\Psi(y,x;b))  \nonumber \\
  - \frac{1}{N} Tr( \Psi(x,y;b) \Psi(y,x;b))))
\eea
that in the large-$N$ limit reduces to:
\bea
0=\int D\mu'  \exp(-\Gamma)
  (Tr(\frac{\delta \Gamma}{\delta \mu'(w)} \Psi(x,x;b))  \nonumber \\
  - \int_{C(x,x)} dy_z \frac{1}{2} \bar{\partial}^{-1}(w-y)Tr( \Psi(x,y;b))
    Tr(\Psi(y,x;b))) 
\eea
where in our notation we have omitted the integrations $D n D \bar n$ since they are irrelevant
in the loop equation. This occurs because the curvature of $B$ depends only on $\mu$.
Gauge invariant functionals of $\mu$ are therefore our analogue of the chiral ring of $ \cal{N}$ $ =1 $
$SUSY$ gauge theories.
Because the trace of an open loop vanishes, the only non-trivial case in Eq.(31) is when
$w$ lies on the loop $C$. In this case
the loop equation can be transformed easily into an equation for $B$ since 
the trace is over the holonomy of a closed loop.
It is clear that the contour integration in the quantum term of the loop equation
includes the pole of the Cauchy
kernel. We need therefore a gauge invariant regularization.
The natural choice consists in analytically continuing the loop equation
from Euclidean to Minkowskian space-time. Thus $z \rightarrow i(x_+ + i \epsilon)$.
This regularization has the great virtue of being manifestly gauge invariant.
In addition this regularization is not loop dependent.
The result of the $i \epsilon$ regularization of the Cauchy kernel is the sum of
two distributions, the principal part plus
a one dimensional delta function: 
\bea
\frac{1}{2}\bar{\partial}^{-1}(w_x -y_x +i\epsilon)= (2 \pi)^{-1} (P(w_x -y_x)^{-1}
- i \pi \delta(w_x -y_x))
\eea
The loop equation thus regularized looks like:
\bea
 0=\int D\mu' \exp(-\Gamma) 
  (Tr(\frac{\delta \Gamma}{\delta \mu'(w)} \Psi(x,x;b)) \nonumber \\
   - \int_{C(x,x)} dy_x(2 \pi )^{-1} (P(w_x -y_x)^{-1}
  - i \pi \delta(w_x -y_x))  \nonumber \\
  \times Tr( \Psi(x,y;b)) Tr(\Psi(y,x;b)))
\eea
Being supported on open loops, the principal part does not contribute and the
loop equation reduces to:
\bea
0=\int D\mu' \exp(-\Gamma)
  (Tr(\frac{\delta \Gamma}{\delta \mu'(w)} \Psi(x,x;b)) \nonumber \\
   +\int_{C(x,x)} dy_{x}
\frac {i}{2}\delta(w_x -y_x) Tr( \Psi(x,y;b)) Tr(\Psi(y,x;b)))
\eea
Taking $w=x$ and using the transformation properties of the holonomy of $b$ 
and of $\mu(x)'$, the preceding equation can be rewritten in terms
of the connection, $B$, and the curvature, $\mu$:
\bea
0=\int D\mu' \exp(-\Gamma)
  (Tr(\frac{\delta \Gamma}{\delta \mu(x)} \Psi(x,x;B)) \nonumber \\
   +\int_{C(x,x)} dy_{x}
\frac {i}{2}\delta(x_x -y_x) Tr( \Psi(x,y;B)) Tr(\Psi(y,x;B)))
\eea
where we have used the condition that the trace of open loops vanishes
to substitute the $b$ holonomy with the $B$ holonomy. \\
We are now ready to construct the quantum holographic effective action
$\Gamma_q$.
On a dense set in the functional integral (in the sense of distributions), the equations:
\bea
F_A-i \Psi^2= \sum_{p} \mu^0_p \delta^{(2)}(x-x_p)- H1\nonumber \\
\bar{\partial}_A \psi= \sum_{p}  n_p \delta^{(2)}(x-x_p)\nonumber \\
\partial_{A} \bar{\psi}=\sum_{p}  \bar{n}_p \delta^{(2)}(x-x_p)
\eea
define an infinite dimensional
twisted local system or, what is the same, a twisted parabolic Higgs bundle on a sphere.
Since the Higgs field acts on the infinite dimensional Hilbert space of a non-commutative $R^2$,
the curvature equation involves a central term, $H$,
equal to the inverse of the parameter of non-commutativity,
$\theta$. This occurs because, once the gauge connection is required to vanish at infinity
up to gauge equivalence, the only term that survives in the curvature at infinity
is the commutator
of the derivatives on the non-commutative $R^2$, that is $H 1$. 
$H 1$ vanishes in the large-$N$ limit. However its trace, i.e. the first Chern class,
need not to vanish in the large-$N$ limit, as we will see momentarily.
The central extension $H 1$ is referred to in this paper as the twist of the local system.
In the case $n=\bar n=0$, that will be the most relevant for us, we may interpret the preceding
equations as vortex equations.
The central extension $H 1$ is related to the non-vanishing of the Higgs field $\Psi$ at infinity,
while the zeroes of the Higgs field are localised at the points at which the hermitean
part of the curvature has delta-like singularities. 
There is a corresponding form of the holographic loop equation in terms of the lattice field of
curvatures of the twisted parabolic Higgs bundles:
\bea
0=\int \prod_q D\mu_q ' D\bar \mu_q ' \exp(-\Gamma)
(Tr(\frac{\delta \Gamma}{ \delta \mu_p}\Psi(x_p,x_p;B))  \nonumber \\
- \int_{C(x_p,x_p)} dy_z \frac{1}{2}\bar{\partial}^{-1}(x_p-y)Tr(\Psi(x_p,y;B))
Tr(\Psi(y,x_p;B)))
\eea
Because of the occurrence of a central extension in the curvature of the non-commutative theory, we need
to modify slightly our formulae for the classical action and for the Chern classes.
In addition we must take into account the parabolic structure, in such a way to define
the parabolic first and second Chern classes \cite{S4}.
Fibrations of parabolic Higgs bundles have been introduced
in the $YM$ functional integral in \cite{MB2} and more recently in \cite{Witt} in $ \cal{N}$ $ = 4 SUSY$
gauge theories.
For twisted parabolic Higgs bundles the functional integral is given by:
\bea
Z=\int \exp(-\frac{N}{2g^2} \sum_{\alpha \neq \beta} \int Tr_f (F_{\alpha \beta}^2)- Tr_f (F_{\alpha \beta})^2 d^4x) DA \nonumber \\
=\int \exp(-\frac{N 8 \pi^2 }{g^2} PC_2-\frac{N}{4g^2} \sum_{\alpha \neq \beta}
 \int Tr_f(F^{-2}_{\alpha \beta})- Tr_f (F^-_{\alpha \beta})^2 d^4x) DA 
\eea
$PC_2$ is the parabolic second Chern class, given by:
\bea
PC_2=\frac{1}{16 \pi^2} \sum_{\alpha \neq \beta} \int Tr_f (F_{\alpha \beta} \tilde F_{\alpha \beta})
-Tr_f (F_{\alpha \beta}) Tr_f(\tilde F_{\alpha \beta}) d^4x \nonumber \\
=C_2+\sum_p (Tr(\lambda_p e_{D_p})-Tr(\lambda_p)Tr(e_{D_p})+\frac{1}{2} (Tr(\lambda^2_p)-Tr(\lambda_p)^2) D^2_p) 
\eea
where $C_2$ is the ordinary second Chern class, $\lambda_p$ the parabolic weight at $p$,
$D^2_p$ the self intersection number at $p$, $e_{D_p}$
the electric flux divided by $2 \pi$ through the dimension two divisor $D_p$ (in fact these fluxes are
referred to as magnetic in
\cite{Witt}, but here we call them electric, since they are dual to the fluxes through the plane of the Wilson
loop that we have referred to as magnetic). 
The parabolic weight is the eigenvalue of the hermitean part of the curvature in the Hitchin
equation, divided by $2 \pi$ and modulo $1$.
$PC_1$ is the parabolic first Chern class, given by:
\bea
PC_1=\frac{1}{4 \pi} \sum_{\alpha \neq \beta}
 \int Tr_f (F_{\alpha \beta}) \tilde \omega_{\alpha \beta} d^4x \nonumber \\
 =C_1+\sum_p Tr(\lambda_p) Deg(D_p)
\eea
where $C_1$ is the ordinary first Chern class and $Deg(D_p)$ the degree of the dimension two
divisor $D_p$. For a detailed explanation see \cite{S4} and in physical language \cite{Witt}. 
In the loop equation there is no contribution from the variation of $PC_2$, since it is a topological
invariant and in particular it vanishes, together with $PC_1$, for stable parabolic Higgs bundles \cite{S4}.
These vanishing constraints must be implemented by delta functions in the functional
integral of the reduced twisted parabolic theory.
Of course $PC_2$ and $PC_1$ contain some four dimensional information on the fibration
of the parabolic Higgs bundles, i.e. a choice of
ordinary Chern classes, electric fluxes, self-intersection numbers and degrees.
These choices introduce a discrete ambiguity and can be justified a posteriori
by the computation of the beta function.
However, these choices may have a natural interpretation
from the point of view of the $EK$ reduction from four to two dimensions.
In particular, to get the correct first coefficient of the beta function, we need a certain
matching between the value of the classical action in the $PC_2=PC_1=0$ sector and the 
number of zero modes that occur in the localisation determinant.
This matching is different, but somehow analogue, to the matching that occurs
for instantons in one-loop perturbation theory. In the latter case the classical action
is $\frac{N (4 \pi)^2 |Q|}{2g^2}$, while the number of zero modes of the operator in the localisation
determinant is $4N|Q|$ (see, for example, \cite{Bian}). In the instanton case, because of the $ASD$ equation $F^-_{\alpha \beta}=0$,
the localisation determinant and the usual one-loop contribution coincide.
In the present case, it turns out that the locus in the moduli space of the twisted parabolic Higgs
bundles, for which the first coefficient of the beta function is reproduced, corresponds to a system
of irreducible twisted Hodge bundles, that in physical terms are vortex equations with
(minus) first Chern class equal to $k$. For such a system the action is $\frac{N (4 \pi)^2 |k|}{2g^2}$, while
it follows by an index theorem that the number of (real) zero modes for fixed parabolic weights is $2N|k|$.
Let us notice that for $|Q|=|k|$ the classical action for instantons and vortices has the same value,
but the number of zero modes for vortices is one half of the number of zero modes for instantons.
Nevertheless we will see
in the next section that the one-loop beta function is the same,
because the spin contribution to the beta function in the instanton case is zero
while it is not so in the vortex case. Therefore we want the condition $PC_2=0$ to imply
the value  $\frac{N (4 \pi)^2 k}{2g^2}$ for the classical action.
We will see at the end of this section how this constraint may be satisfied.
Now, to construct the quantum holographic effective action, we add infinitesimal strips
starting from the loop and ending into the parabolic points of the two regions in which the sphere is divided by
the loop. Then we map conformally each region to a cuspidal fundamental domain over which
the quantum term vanishes because of the zig-zag symmetry (the loop backtracks in a neighbourhood of the cusps).
The strips occur in pairs, therefore each chart of the hologram has the same number of cusps and
the same must hold for the parabolic points of the charts in the original space time.
Mathematically the family of arcs on the hologram belongs to the arc complex of a Riemann surface \cite{Pen}.
In fact, the pairing of the arcs that intersect the loop in our approach to the loop equation,
matches exactly the way weighted arc families are composed on a Riemann surfaces \cite{Pen}.
In the loop equation there is no logarithmic cuspidal anomaly,
because, when the arcs ending into a cusps are parallel, the cuspidal anomaly
becomes linearly divergent, rather than logarithmically divergent, and thus it mixes with
the usual linearly divergent contribution proportional to the perimeter,
due to short distance Coulomb-like behaviour \cite{Gross2}.
Since we can choose arbitrarily a parabolic point on the loop,
the loop has a cuspidal image in the point at infinity.
We should recall the reader that there exist two different mathematical versions of the 
uniformization theory \cite{Pen}: the one in the hyperbolic setting and the one in the conformal setting.
Of course they are equivalent topologically, but not metrically.
The hyperbolic setting is reminiscent of the $AdS$ correspondence \cite{Mal}, but the version that works
at the level of loop equation for our approach is the conformal setting, since we want that the scaling
factor of the metric be induced by a conformal diffeomorphism.  
Quadratic differentials \cite{Pen1} can be used to construct the uniformization
map to the cuspidal fundamental domain.
The basic relation between quadratic differentials, $q$, and the uniformization map $t$ is:
\bea
\frac{\partial t}{\partial z}= \sqrt q
\eea
We need therefore the standard form of a quadratic differential
near a cusp:
\bea
\frac{\partial t}{\partial z}=  \frac{L}{2 \pi i z}
\eea
where $L$ is the length of the horocycle arc around the cusp.
Since this expression is infinite at the cusps it must be regularized and suitably
interpreted.
In particular it depends crucially on what the cutoff is on the fundamental
domain near the cusps. 
We must distinguish the $d=2$ from the $d=4$ case.
In the $d=2$ case, the theory is invariant under an area preserving
diffeomorphism, 
and thus it is not restrictive to consider a circular loop of area equal to the area of the region inside
the loop. In this case we have essentially a circle
that is mapped to the circle at infinity. This is a cylinder, i.e. a punctured disk,
that is mapped by the uniformization map to a strip in the upper-half plane.
In this case the uniformization map is:
\bea
t= \frac{L}{2 \pi i }log(z)
\eea
Thus we get:
\bea
|\frac{\partial t}{\partial z}|(p)^2= \frac{R^2}{a^2}=\frac{A}{ \pi a^2} = N_D
\eea
where $R$ is the radius and $A$ the area of the disk, while $a$ is the radius
of a little disk around the puncture. Thus $N_D$ is the number of lattice
points inside the disk. In the $d=4$ case, the theory is not invariant under an area preserving
diffeomorphism, however, as far as we are interested in the local approximation
for the quantum effective action, the only thing that matters
is how to interpret the quadratic differentials near the cusps, and it is natural
to maintain the $d=2$ interpretation. More intrinsically, in the computation of the quark-antiquark
potential (that we do not perform here), we would be interested in a very long rectangular Wilson loop.
The two long parallel sides of the loop would then be, on a (large) sphere, two circles of very small curvature,
in such a way that the two dimensional interpretation holds literally. 
At global level, in the conformal setting, the most suggestive representation of the hologram
is as a Mandelstam graph, in which the infinitesimal strips are strings ending into the cusps
(see, for example, \cite{Pen, Man}). 
Finally the holographic loop equation on the hologram reads:
\bea
0=\tau(\frac{\delta \Gamma_{q}}{\delta \mu_{p}}\Psi(x_p,x_p;B)) 
\eea
($\tau$ denotes the combination of the colour trace with the v.e.v.) 
that is implied by the critical equation:
\bea
\frac{\delta \Gamma_{q}}{\delta \mu_{p}}=0
\eea
which we refer to as the master equation. \\
We are now ready to construct the quantum holographic effective action.
We denote by $\Gamma^{\infty}_q$ and $\Gamma^{0}_q$ the quantum effective actions
on the corresponding charts of the hologram.
They are obtained in the following way. It is repeated
the construction of $\Gamma$ in each chart of the hologram.
On the hologram the axial gauge and the gauge $\mu_p^-=0$ at the cusps are chosen.
The label $^-$ for $\mu_{p}$ means here lower
triangular part, excluding the diagonal, while the label $^+$ for $\mu_{p}$
means here upper triangular part including the diagonal.
In this gauge $log|\frac{D \mu}{D \mu'}|=0$ because the gauge conditions $\mu_p^-=0$ and ${\mu'}_p^-=0$ 
can be imposed at the same time,
by means of gauge transformations respectively unitary and in the complexification of the gauge group,
and the resulting fields may differ only by transformations that are upper triangular,
thus giving trivial contribution to the Jacobian.
Let us observe also that $Det(ad\mu^+_p)|_{\mu_p^-=0}$ reduces to
the Vandermonde determinant of the eigenvalues of $\mu_p$ and as such
can be written in any gauge.
Thus in the $d=2$ theory, $\Gamma_q$ on each chart of the hologram is the 
classical action on the hologram minus the logarithm the Vandermonde
determinant of the eigenvalues. 
Since the $d=2$ theory is not conformal invariant the classical action on the hologram
differs by a conformal factor of the metric from the classical action on space-time,
as we will see at the beginning of sect.3 . Finally to get the effective action on the entire hologram
we multiply the contributions from each chart and sum over the discrete set of gauge orbits that
leaves invariant the holonomy on the boundary of each chart.\\
In fact we should take into account some extra Jacobians that arise by imposing that the holonomy
on each chart has a fixed value and the $FP$ that arises choosing a gauge in which
the holonomy is actually diagonal. It is possible to show, however \cite{Cas}, that the product of these
Jacobians cancels exactly.
In $d=4$, on the hologram, $\Gamma_q$ is the same as $\Gamma$ on space-time, up to the conformal anomaly,
because the hologram is a conformal image of the punctured space-time
in each chart. As in $d=2$ we impose an axial gauge in a direction orthogonal to the line of cusps.
Using the extended gauge symmetry across the cusps, we set $\mu$ in upper
triangular form and we add to $\Gamma$ minus the logarithm of the corresponding $FP$ determinant,
that in this case too is the Vandermonde determinant. This completes the construction of $\Gamma_q$
in each chart. \\
More explicitly:
\bea
\Gamma_{q}= \Gamma |_{A_y=0} -\sum_p log Det(ad\mu^+_p)|_{\mu_p^-=0}
+ Conformal Anomaly
\eea  
The first term, $\Gamma|_{A_y=0}= (\frac{N}{2g^2}S_{YM}
+\frac{1}{2}logDet'(-\Delta_A \delta_{\alpha \beta}
+ D_{\mu} D_{\nu} +i ad_{F^-_{\alpha \beta}}))|_{A_y=0}$, is the classical holographic action 
associated to the reduced non-commutative theory in the axial gauge on the space-time.
For the purpose of computing the beta function, $\Gamma$ can be computed in any gauge,
provided we add the logarithm of the corresponding Faddeev Popov determinant.
Finally, we sum over a discrete set of gauge orbits, that leaves invariant the holonomy of $B$,
and also over all sectors of $Z_N$ flux, since we are in fact computing
a Wilson loop in the adjoint representation.
This completes the construction of $\Gamma_q$ on the entire hologram in $d=4$.
The $d=4$ theory has some peculiarity, since we are representing
an adjoint Wilson loop as the product of the fundamental one and its conjugate in the large-$N$ limit.
The orbifold and orientifold case correspond to merge in a different way the
contributions from the two charts of the hologram.
We can explain the two constructions in terms of two different resolutions of identity in the functional 
integral.
In the orbifold case we merge together the chart at zero and the chart at infinity 
that have as boundary a Wilson loop in the same representation.
We get therefore for the resolution of identity of the reduced orbifold theory:
\bea
1=\sum_k \int \delta(F_B - \sum_p (\mu_{0p} \delta^{(2)}_{0p}+\mu_{\infty p} \delta^{(2)}_{\infty p})+
H_k 1)   \prod _p D \mu_p
\times \nonumber \\
\int \delta(\bar F_B - \sum_p (\bar \mu_{0p} \delta^{(2)}_{0p}+\bar \mu_{\infty p} \delta^{(2)}_{\infty
p})- H_k 1)   \prod
_p D \bar \mu_p
\eea
On the orbifold the partition function, for a fixed $Z_N$ flux, factorizes
into the contributions
of the fundamental and the conjugate representation. The complete partition
function is obtained summing over sectors
of equal $Z_N$ flux. The gauge group of the theory is, in an effective way, $\frac{SU(N)_f \times  SU(N)_{\bar f} }{ Z_{N}}$.
In the orientifold case, instead, we merge the two charts at infinity and the two charts at zero, that have 
as boundary a Wilson loop in the fundamental and in the conjugate representation.
The merging is possible since the v.e.v. of the Wilson loop are the same for both the representations
\cite {Ve2} and thus we assume that the
eigenvalues in the two different representation can be identified up to gauge equivalence.
In fact, since the v.e.v of a Wilson loop is real, we expect the eigenvalues 
to occur in pairs, with opposite sign in the exponent. The conjugate representation
will change just the sings, allowing the identification by re-ordering.
The assumption that we make is equivalent to requiring that charge conjugation is unbroken
in the large-$N$ limit and that acts by gauge transformations on the eigenvalues of the Wilson loop.
We get therefore for the resolution of identity of the reduced orientifold theory:
\bea
1=\sum_k \int \delta(F_B - \sum_p \mu_{0p} \delta^{(2)}_{0p}+ H_k 1) \times \nonumber \\
\delta(\bar F_B - \sum_p \bar \mu_{0p} \delta^{(2)}_{0p} - H_k 1) \prod _p D \mu_{0p}
D \bar \mu_{0p}
\times \nonumber \\
\int \delta(F_B - \sum_p  \mu_{\infty p} \delta^{(2)}_{\infty p}+ H_k 1) \times \nonumber \\
\delta(\bar F_B - \sum_p \bar \mu_{\infty p} \delta^{(2)}_{\infty p}- H_k 1)
\prod_p D \mu_{\infty p}
D \bar \mu_{\infty p}
\eea
On the orientifold the partition function, for a fixed $Z_N$ flux, factorizes into the contributions
of two non-orientable surfaces
that are obtained doubling the chart at zero and the chart at infinity by gluing
through an orientifold plane over which the Wilson loop lies. The complete partition function
is obtained summing over sectors
of equal $H_k$ field. A subtle point arises about the first Chern class of the parabolic bundles on the two
holograms in the orientifold theory. Cutting and gluing changes the flux, thus to get the same flux,
we perform a dilatation whose effect is taken into account by the conformal anomaly.
The gauge group of our orientifold theory is then, in an effective way, a $SU(N)$ diagonally embedded into 
$\frac{SU(N)_f \times  SU(N)_{\bar f} }{ Z_{N}}$ on the (non-orientable) double of space-time.
Interestingly, in the orientifold case, we are somehow separating the infrared from
the ultraviolet, since we are taking different continuum limits on the two holograms,
thanks to the fact that the lattice spacings are different, as we will see shortly. 
In this respect it would be interesting to study the action of a duality
transformation on the partition function, although it will not be considered here.
In the resolution of identity in Eq.(48,49) we have omitted the sum over the shifts of the parabolic weights,
but we will display them in the quantum holographic effective action.
The allowed shifts are the ones that leave
invariant the holonomy of the loop and we leave them undetermined in this paper.
However the shift ensemble is non-void, since it has to contain at least the shifts in the curvature
associated to the action of the $Z_N$
group of large gauge transformations. 
Finally, we should find out which is the value of the classical action implied
by the stability constraints, $PC_2=PC_1=0$, on the hologram in the orientifold case.
As we will see in the next section, we would like to have $ \sum_p Tr(\lambda^2 _{\infty p}+ c.c.) -2k=0$
to get the correct beta function. This implies a number
of choices in Eq.(39). If we set, as it seems natural, $D^2_{\infty p}=1$ for all $p$, then we need
$C_2- \frac{1}{2} C_1^2-k tr(e_{D_{\infty p}})+k tr(\bar e_{D_{\infty p}})=0$ while $ke_{D_{\infty p}}-k \bar
e_{D_{\infty p}}+\frac{1}{2} \sum_p Tr(\lambda^2 _{\infty p}+ c.c. )=0$. Thus $e_{D_{\infty p}}=-\frac{1}{2}$
and $ \bar e_{D_{\infty p}}=\frac{1}{2}$ for all $p$. $C_1=-k$ and $\bar C_1= k$ by the stability
condition $PC_1+ P \bar C_1= 0$ on the orientifold, with the natural choice $Deg(D_{\infty p})=1$. 
Thus the $EK$ reduction in the plane transverse to the Wilson loop has to be made in presence
of an electric flux with $e_{D_{\infty p}}=-\frac{1}{2}$ and $\bar e_{D_{\infty p}}=\frac{1}{2}$,
that, on the (non-orientable) double cover, still satisfies the quantisation condition 
$k(- \frac{1}{2})+ (-k)\frac{1}{2}=integer$.
We will see in the next section that these constraints are actually non-void for a vortex system.
Had we employed the hermitean resolution of identity, we would have needed
$e_{D_{\infty p}}=-\bar e_{D_{\infty p}}=-1$, in order to get the correct normalisation of the classical action, and thus 
the correct beta function. Yet, we have been unable to show, in this case, that the associated constraint
is non-void for a vortex system.

\section{The large-$N$ exact beta function for the Wilsonean and the canonical coupling}

For the reader convenience, as an exercise before considering explicit formulae for the quantum
effective action in $d=4$, we write the formulae for the $d=2$ case. 
$\Gamma_q$ in $d=2$, in the fundamental representation, reads:
\bea
\exp(-\Gamma_{q}) 
= \sum_{k_0,k_{\infty}} \exp(-\frac{N}{ g^2 \tilde a^2} \sum_i \sum_p |\frac{\partial t}{\partial z}(p)|_0 ^2
 (h+k_0)^{i 2}_p  \nonumber \\
 +\sum_{i > j} \sum_p log(h^{i}_p -h^{j}_p +k_{p0} ^{i } -k_{p0} ^{j }))  \nonumber \\ 
\times
\exp(-\frac{N}{ g^2 a^2} \sum_i \sum_p |\frac{\partial t}{\partial z}(p)|_{\infty} ^2
 (h+k_{\infty})^{i 2}_p  \nonumber \\
 +\sum_{i > j} \sum_p log(h^{i}_p -h^{j}_p +k_{p \infty} ^{i }-k_{p \infty} ^{j }))
\eea
where  $h_p$ is the lattice field of the eigenvalues of the curvature.
We have set $a=\frac{2 \pi}{\Lambda}$, 
with $a$ the lattice spacing
corresponding to the cutoff $\Lambda$ of the theory, that arises from the product of delta
functions at the same point in the classical action, and analogously for $ \tilde a$. 
Since the number of points, $N_D$, is the same in each chart by construction, the only way to define
different areas is to choose different values of the lattice spacing in each chart , $a$ and $\tilde a$.
The shifts of the eigenvalues of the curvature, $k_0$ and $k_{\infty}$, are chosen in such a way to leave invariant
the Wilson loop. Therefore they satisfy the conditions:
\bea
\sum_p k_{p 0} ^{j } = 2 \pi \times integer \nonumber \\
\sum_p k_{p \infty} ^{j } = 2 \pi \times integer
\eea
It should be noticed that $\Gamma_{q}$ is expressed
as a functional of the curvature on the hologram. This
involves a change of the metric in the classical action, since the classical action is not
conformally invariant. Using our interpretation of the regularized quadratic differentials, we get:
\bea
\exp(-\Gamma_{q}) 
= \sum_{k_0,k_{\infty}} \exp(-\frac{N N_D}{ g^2 \tilde a^2} \sum_i \sum_p 
 (h+k_0)^{i 2}_p \nonumber \\
 +\sum_{i >j} \sum_p log(h^{i}_p -h^{j}_p +k_{p0} ^{i } -k_{p0} ^{j }))  \nonumber \\ 
 \times
\exp(-\frac{N N_D}{ g^2 a^2} \sum_i \sum_p 
 (h+k_{\infty})^{i 2}_p \nonumber \\
+\sum_{i > j} \sum_p log(h^{i}_p -h^{j}_p +k_{p\infty} ^{i } -k_{p\infty} ^{j }))
\eea
Assuming translational invariance in each of the two charts of the hologram:
\bea
k_{p0} ^{j }  =k_{0} ^{j }= \frac{2 \pi}{N_D} \times integer \nonumber \\
k_{p\infty} ^{j } =k_{\infty} ^{j } = \frac{2 \pi}{N_{D}} \times integer
\eea
Thus, the  quantum effective action reduces to:
\bea
\exp(-\Gamma_{q}) 
= \prod_p \sum_{k_0 k_{\infty}}\exp(-\frac{N N_D}{ g^2 \tilde a^2} \sum_i  
 (h+k_0)^{i 2} + 
  \sum_{i > j}  log(h^{i} -h^{j} +k_0 ^{i } -k_0 ^{j }))  \nonumber \\ 
 \times
\exp(-\frac{N N_D}{ g^2 a^2} \sum_i  
 (h+k_{\infty})^{i 2} +
  \sum_{i > j}  log(h^{i} -h^{j}+k_{\infty} ^{i } -k_{\infty} ^{j }))
\eea
where the $h^i$ are determined as follows.
We can express the effective action in terms of the eigenvalues of the holonomy, $\exp (i \lambda^i)$,
instead of the eigenvalues of the curvature of the local system.
We have the relations:
\bea
(\lambda+2 \pi \times integer)^i=N_D (h+k_0)^{i}= N_{D}(h+k_{\infty})^{i}
\eea
obtained requiring that the holonomy at the boundary of each chart
is the same and assuming translational invariance on each chart. 
The quantum effective action then satisfies:
\bea
\exp(-\frac{\Gamma_{q}}{N_D}) = 
 \sum_{m_0, m_{\infty}}\exp(-\frac{N}{ g^2 N_D \tilde a^2} \sum_i 
 (\lambda+2 \pi m_0)^{i 2}  \nonumber \\
 + \sum_{i > j}  log(\lambda ^{i} - \lambda ^{j} +2 \pi m_0 ^{i } -2 \pi m_0 ^{j }))  \nonumber \\ 
 \times
\exp(-\frac{N}{ g^2 N_{D} a^2} \sum_i
 (\lambda+2 \pi m_{\infty})^{i 2}  \nonumber \\
 + \sum_{i > j}  log(\lambda ^{i} -\lambda ^{j}+2 \pi m_{\infty} ^{i } -2 \pi m_{\infty} ^{j }))
\eea
where now the sum over $m_0, m_{\infty}$ is on integers.
$\Gamma_{q}$ coincides exactly, up to a factor of $N_D$, irrelevant in the loop equation,
with the quantum effective action for the eigenvalues
of a Wilson loop on a sphere obtained by functional methods \cite{Cas}, provided
we identify $A_0=N_D a^2$ and $A_{\infty}=N_{D} \tilde a^2$, where $A_0$ and $A_{\infty}$ are the
areas of the two charts in which the sphere is divided by the Wilson loop. \\
We now pass to the $d=4$ case.
Following \cite{MB1}, it is convenient to perform the computation of the divergent parts of
$\Gamma_q$ in an indirect way, by means of a term by term comparison with the usual one-loop
perturbative contribution to the effective action. 
For this purpose, let us recall the structure of one-loop perturbative
corrections to the classical action, in the Feynman gauge: 
\bea
\int Dc DA_{\alpha} \exp(-\frac{N}{2 g^2} \int d^4 x Tr(c^2)) \exp(-\frac{N}{2 g^2} S_{YM}) \delta(D_{\alpha} 
\delta A_{\alpha}-c) \Delta_{FP}= \nonumber \\
=\exp(-\frac{N}{2 g^2}S_{YM})Det^{-\frac{1}{2}}(-\Delta_A \delta_{\alpha \beta}+i 2 ad_{F_{\alpha \beta}})
 Det(-\Delta_A) 
\eea
where we have inserted in the functional integral the gauge-fixing condition and the corresponding
Faddeev-Popov determinant and, by an abuse of notation, we have denoted with $A$
the classical background field in the right hand side of Eq.(57). It follows that
the perturbative one-loop effective action, in the Feynman gauge, is given by:
\bea
\Gamma_{one-loop}= \frac {N}{2 g^2}S_{YM}+\frac {1}{2} 
logDet(-\Delta_A \delta_{\alpha \beta}+i2ad_{F_{\alpha \beta}})- logDet(-\Delta_A )
\eea
The perturbative computation of the one-loop beta function \cite{AF,AF1} is the result of
two contributions, that are independent within logarithmic accuracy \cite{Pol3}.
The orbital contribution gives origin to diamagnetism and to a positive 
term in the beta function:
\bea
-log(Det^{-\frac{1}{2}}(-\Delta_A \delta_{\alpha \beta}) Det(-\Delta_A))=
log Det(-\Delta_A)= \nonumber \\
= \frac{1}{3} \frac{N}{(4 \pi)^2}log(\frac {\Lambda}{\mu}) \sum_{\alpha \ne \beta} \int d^4x Tr_f
(F_{\alpha \beta})^2
\eea
where it should be noticed the cancellation of two of the four polarisations between
the first factor and the Faddev-Popov determinant.
The spin contribution gives origin to paramagnetism and to an
overwhelming negative term in the beta function \cite{Pol3}:
\bea
\frac{1}{4} \sum_{\alpha \ne \beta} Tr(i2 ad_{F_{\alpha \beta}} (-\Delta_A )^{-1}
i2 ad_{F_{\alpha \beta}} (-\Delta_A )^{-1})= \nonumber \\
= - 4 \frac{N}{(4 \pi)^2} log(\frac {\Lambda}{\mu})
  \sum_{\alpha \ne \beta}\int d^4x Tr_f (F_{\alpha \beta})^2
\eea
Hence the complete result for the divergent part of $\Gamma_{one-loop}$ is:
\bea
(\frac {N}{2 g^2}- \frac{11}{3} \frac{N}{(4 \pi)^2}log(\frac {\Lambda}{\mu})) 
 \sum_{\alpha \ne \beta} \int d^4x Tr_f (F_{\alpha \beta})^2
\eea
In the sector with $PC_2=0$, which is the one of stable parabolic Higgs bundles that are of interest for us,
this reduces to:
\bea
(\frac {N}{2 g^2}- \frac{11}{3} \frac{N}{(4 \pi)^2}log(\frac {\Lambda}{\mu})) 
 \sum_{\alpha \ne \beta} \int d^4x Tr_f (2 (\frac{1}{2} F^-_{\alpha \beta})^2)
\eea
Now, the orbital contribution, from the localisation determinant in
$\Gamma_q$, is the same as the one-loop perturbative one,
and thus, in the $PC_2=0$ sector, reduces to:
\bea
\frac{1}{3} \frac{N}{(4 \pi)^2}log(\frac {\Lambda}{\mu}) \sum_{\alpha \ne \beta} \int d^4x 
Tr_f (2(\frac{1}{2}F^-_{\alpha \beta})^2)
\eea
On the contrary, the spin contribution, from the localisation determinant in $\Gamma_q$,
involves only the anti-selfdual part of the curvature, instead of
the complete curvature and, in every sector, is equal to:
\bea
- 4 \frac{N}{(4 \pi)^2} log(\frac {\Lambda}{\mu})
  \sum_{\alpha \ne \beta}\int d^4x Tr_f (\frac{1}{2} F^{-}_{\alpha \beta})^2
  \eea
Hence, the spin contribution in $\Gamma_q$, from the localisation determinant,
is only one half of the spin
contribution in perturbation theory in the $PC_2=0$ sector.
Thus, the orbital and spin contributions in $\Gamma_q$, from the localisation determinant,
in the $PC_2$ sector, sum up to: 
\bea
(\frac {N}{2g^2}-(2-\frac {1}{3})\frac{N}{(4 \pi)^2} log(\frac {\Lambda}{\mu}))
\sum_{\alpha \ne \beta} \int d^4x 
Tr_f (2(\frac{1}{2}F^-_{\alpha \beta})^2) \nonumber \\
=(\frac {N}{2g^2}-\frac {5}{3}\frac{N}{(4 \pi)^2} log(\frac {\Lambda}{\mu}))
\sum_{\alpha \ne \beta} \int d^4x Tr_f (2(\frac{1}{2}F^-_{\alpha \beta})^2)\nonumber \\
=\frac {N}{2g^2} Z^{-1} \sum_{\alpha \ne \beta} \int d^4x Tr_f (2(\frac{1}{2}F^-_{\alpha \beta})^2)
\eea
where $Z^{-1}$ is given by:
\bea
Z^{-1}=1-\frac{10}{3} \frac{1}{(4 \pi)^2} g^2 log(\frac {\Lambda}{\mu})
\eea
Therefore the localisation determinant alone does not reproduce the exact
one-loop beta function in $\Gamma_q$.
The only source of new divergences in $\Gamma_q$ can be normalizable zero modes,
arising in the integration on the gauge connection of the delta function in Eq.(18).
Until now we have tacitly assumed that no normalizable zero mode occurs, but we have just
understood that in order to reproduce the exact one-loop beta function they 
are in fact necessary. 
The dimension of the space of zero modes is equal to the dimension of the moduli space 
of those deformations
of the parabolic Higgs bundles that leave invariant the functional:
\bea
\int d^2x Tr_f (F^{-}_{\alpha \beta}-\mu^{-}_{\alpha \beta})^2
\eea
that occurs in the definition of the delta function, as the limit of an exponential of a quadratic form,
provided the corresponding zero modes are normalizable.
Thus, formally, these moduli are associated in general to adjoint orbits.
However, for a Higgs bundle, the deformations associated to adjoint orbits
do not lead in general to normalizable the zero modes, since,
in the tangent space to adjoint orbits, delta-like singularities occur,
for fixed $\lambda_p, n_p$:
\bea
F_A-i \Psi^2= \sum_{p} 2 \pi  g_p \lambda _p g^{-1}_p \delta^{(2)}(x-x_p)- H 1 \nonumber \\
\bar{\partial}_A \psi= \sum_{p} n_p \delta^{(2)}(x-x_p) \nonumber \\
\partial_{A} \bar{\psi}=\sum_{p} \bar n_p  \delta^{(2)}(x-x_p) 
\eea
In Eq.(68) we have rescaled the eigenvalues of the adjoint orbits by a factor of $2 \pi$, in such a way
that they coincide now with the parabolic weights modulo $1$.
Eq.(68) describes a Kahler quotient by the compact gauge group, in which the dimension of the moduli 
space of solutions, for fixed $\lambda_p, n_p$, is the dimension of the adjoint orbit.
This is the dimension of the space of formal zero modes of the quadratic form in Eq.(67). 
The complex dimension of a $SU(N)$ adjoint orbit is
$\frac{1}{2} (N^2-\sum_i m^{i2})$, where $m^{i2}$ are the multiplicities of the eigenvalues
of the adjoint orbit. Thus, if all the eigenvalues are different, the complex dimension is $\frac{1}{2}(N^2-N)$.
However, in general these zero modes are not normalizable, unless the metric is changed and adapted to
the singularities of the connection determined by the adjoint orbits \cite{Biq}.
(On the hologram) we can gauge transform Eq.(68) to:
\bea
F_A-i \Psi^2= \sum_{p} 2 \pi  \lambda _p  \delta^{(2)}(x-x_p)- H 1 \nonumber \\
\bar{\partial}_A \psi= \sum_{p} g^{-1}_p n_p g_p\delta^{(2)}(x-x_p) \nonumber \\
\partial_{A} \bar{\psi}=\sum_{p}  g_p\bar n_p g^{-1}_p \delta^{(2)}(x-x_p) 
\eea
Now, there is a special locus in the moduli space of the
Higgs system, for which
the Higgs field is regular and thus normalizable zero modes may exist in the tangent space.
This locus corresponds to set $n_p=\bar n_p=0$, to get a kind
of infinite dimensional vortex equation \cite{Qui}:
\bea
F_A-i \Psi^2= \sum_{p} 2 \pi  \lambda _p  \delta^{(2)}(x-x_p)- H 1 \nonumber \\
\bar{\partial}_A \psi= 0 \nonumber \\
\partial_{A} \bar{\psi}=0
\eea
In fact this locus is favoured by the master equation in the local approximation,
since the Vandermonde determinant creates repulsion between the eigenvalues of $\mu$,
while the other moduli are suppressed by the classical action. 
On the hologram there are not anymore adjoint orbits.
However, we can determine the dimension of the moduli space of normalizable zero modes as well, using an
index theorem
that counts the number of holomorphic sections of the holomorphic bundle:
\bea
\bar{\partial}_{A} \psi= 0
\eea
The complex dimension of this space is $N|k|$ \cite{index}, where $k$ is minus the first Chern class, that, for 
stable parabolic Higgs bundles
of parabolic degree zero, is given by:
\bea
k=\frac{1}{2 \pi} \int Tr(H_k) d^2x  \nonumber \\
=\sum_p Tr(\lambda_p mod1) \nonumber \\
=-C_1
\eea
Of course the two ways of counting must coincide, and in fact they do, provided singularities
in the moduli of vortex configurations are suitably interpreted.
It is easy to see directly in some special cases how the two ways of counting coincide, provided we take
into account the restrictions of rational type that arise on the eigenvalues, due to the fact that
the singularities of $A$ can only arise from the zeroes of $\psi$.
For one $SU(N)$ vortex the eigenvalues of the curvature are
$\frac{2 \pi}{N}...\frac{2 \pi(1-N)}{N}$. Therefore the complex dimension of the orbit is $\frac{1}{2}(N^2-1-(N-1)^2)=N$
and $k=1$, since all the parabolic weights, being the eigenvalues divided by $2 \pi$ modulo $1$,
are $\frac{1}{N}$.
However, if the eigenvalues are $\frac{2 \pi k}{N}...\frac{2 \pi(1-N)k}{N}$ the counting of the holomorphic 
sections and the dimension of the orbit do not seem to agree. Yet, from the holomorphic point
of view, this configuration describes in fact $k$ vortices colliding at the same point.
Vortices of this kind may occur at strong coupling, where we do not expect
a scaling behaviour.
Let us consider now the more complicated situation of a generic adjoint orbit, that we expect to be relevant
in the large-$N$ limit.
If we where in finite dimension, because Eq.(71) is left invariant under constant
rescaling of $\psi$, it would describe bundles of Hodge type, that are indeed the fixed points of the 
rescaling action from the holomorphic point of view \cite{Hit,S1,S2}.
Thus $\psi$ would be nilpotent \cite{Hit,S1,S2}.
In Eq.(69), instead, Hodge bundles arise as fixed points of the circle action, $\exp(i \alpha)$,
on $\psi$, that must act by gauge transformations \cite{Mu}. 
Thus, $A$ decomposes as a sum of irreducible representations of dimensions equal to the multiplicity
of the parabolic weights, that therefore turn out to be rational.
Yet, $\psi$ may still stay irreducible, provided all the blocks in $A$
have the same dimension \cite{Witt2}, for example are valued in $SU(m)$. Thus $N=mN'$.
In each $SU(m)$ block we put one vortex. Thus $k=N'$, the number
of $SU(m)$ blocks, and the complex dimension of zero modes associated to this orbit is $NN'$.
Of course, to get the total number of zero modes, we have to sum over the dimensions of all
the non-trivial orbits. It is easy to see that, in the large $m$ limit, the contribution
of each vortex of this type to the classical action is 
$\frac{N (2 \pi)^2 N'}{ g^2}$ and therefore the total contribution of vortices
in the fundamental representation is $\frac{N (2 \pi)^2 k}{ g^2}$.
We are ready now to continue our computation of the beta function. 
We may think that, somehow conserving our difficulties,
we have traded the integration over the order of $N^2$ degrees of freedom 
in the curvature of the original theory into the integration over the (eventually) same order of 
$N^2$ degrees of freedom of the normalizable zero modes on the hologram.
Yet, we know from the loop equation, that in the second case the resulting theory must be classical,
in the sense that on the hologram all the quantum corrections are already
included in the quantum holographic effective action.
This is precisely what happens. Because of the normalizable zero modes, $\Gamma_q$
gets the missing contributions, in order to complete the correct one-loop beta function.
In the orientifold case the hologram consists of two disconnected surfaces, the hologram at
zero and the hologram at infinity. We start computing the contribution of the hologram at infinity.
Because of the constraint $PC_2=PC_1=0$, to hold in the large-$N$ limit, and our choices
for the fibration of the Higgs bundles,
the classical action on the hologram at infinity is:
\bea
\exp(-\frac{ 2Nk (2 \pi)^2 }{g^2})
\eea
because we get $k$ vortices in the fundamental representation and $k$ vortices
in the conjugate representation from the doubling of the chart at infinity in the orientifold. 
The localisation determinant renormalises the classical action according to Eq.(65,66):
\bea
\exp(-\frac{2Nk Z^{-1}(2 \pi)^2}{g^2})
\eea
The contribution of zero modes is:
\bea
\Lambda^{2Nk}=(\frac {2 \pi}{\tilde a})^{2Nk}
\eea
Indeed $k$ vortices in the fundamental representation contribute $Nk$ holomorphic
zero modes, that are paired with the $Nk$ anti-holomorphic ones from the doubling with the conjugate
representation. This matching between holomorphic and anti-holomorphic zero modes
follows from the holomorphic resolution of identity employed in Eq.(21) and the pairing with
the corresponding anti-holomorphic contribution in the conjugate representation
in the resolution of identity of the orientifold construction, Eq.(49).    
This pairing on the orientifold resembles the absence of the twisted sector
in the field theoretical orientifold \cite{A1,A2,Ve2}.
Combining Eq.(74,75), we get for the Wilsonean coupling constant:
\bea
\frac{(4\pi)^2 Nk}{2 g^2_W(\mu)} 
 =(4\pi)^2Nk( \frac{1}{2 g^2_W(\Lambda)}- \frac{1}{(4\pi)^2} (2+\frac{5}{3}) 
log (\frac{\Lambda}{\mu}))
\eea
provided there are no other divergent contributions from the hologram at zero,
as we will show in the following.
From a direct computation we get that the classical action on the hologram at zero
scales as $\frac{\tilde a^2}{a^2}$. This is due to the normalisation condition
$\frac{\pi}{N_2 H \tilde a^2}=1$ on the hologram at infinity that implies for the hologram
at zero the normalisation $\frac{\pi}{N_2 H a^2}=\frac{\tilde a^2}{a^2}$.
Then on the hologram at infinity we take the continuum limit $\tilde a \rightarrow 0$ with
$A_{\infty}=N_D \tilde a^2=const$ and small, while on the hologram at zero we keep $a=const$ with
$A_{0}=N_D  a^2 \rightarrow \infty $. This means that on the hologram at infinity we take
the continuum limit at fixed small area, while on the hologram at zero we take the termodinamic limit
at fixed lattice spacing. In this way we send the area of the universe to infinity,
while keeping the area of the loop finite and small. There might be other ways of taking the continuum limit,
however, we will not discuss them in this paper. Any possible ultraviolet quantum contribution
is finite in the hologram at zero, since we have kept $a$ fixed. 
Thus, in the limit $\tilde a \rightarrow 0, a=const$, we conclude that the Wilsonean coupling constant
is exactly one loop. 
However, this will not be the case for the canonical coupling,
as we will see momentarily. 
We start with a preamble, recalling how the difference
between the Wilsonean and the canonical beta function can be understood
in terms of a rescaling anomaly in the functional integral, in the
$ \cal{N} $ $ = 1$ super-symmetric case, following \cite{HM}.
The canonical coupling constant can be related to the Wilsonean one
taking into account an anomalous Jacobian that occurs in the
functional integral:
\bea
Z= \int \exp(-\frac{N}{2g_{W}^2}S_{YM}(A)) DA \nonumber \\
=\int \exp(-\frac{N}{2g_{W}^2}S_{YM}(g_c A_c)) \frac{D(g_c A_c)}{DA_c} DA_c\nonumber \\
=\int \exp(-\frac{N}{2g_{W}^2}S_{YM}(g_c A_c)+ log \frac{D(g_c A_c)}{DA_c}) DA_c \nonumber \\
=\int \exp(-\frac{N}{2g_{c}^2}S_{YM}(g_c A_c)) DA_c 
\eea
From this relation it follows that:
\bea
\frac{N}{2g_{c}^2} =  \frac{N}{2g_{W}^2} -   S^{-1}_{YM}(g_c A_c) log \frac{D(g_c A_c)}{DA_c}
\eea
We now pass to the large-$N$ $QCD_4$ case.
Since we have completed the construction of the
quantum effective action, $\Gamma_q$, in terms of the eigenvalues of the
curvature of a twisted local system, the computation of the canonical beta function
is by now most easily performed.
We should notice that, since our effective action involves only the $N$
eigenvalues, the rescaling anomaly that arises from the integration measure for
them, whatever it is, is sub-leading
in $\frac{1}{N}$. Thus, if there is at all a rescaling anomaly, it must
be already contained in the functional form of the quantum effective action.
Here are the expressions for the local part of the effective action, up to the conformal anomaly and finite
terms, for
the hologram at infinity and at zero respectively:
\bea
\exp(-\Gamma^{k,\infty}_{q})= \sum_e \delta(\sum_{i,p}(\lambda^i_p+e^i_p) -k)
\delta(\sum_{i,p}(\bar \lambda^i_p+ \bar e^i_p) +k) \nonumber \\
\delta(\sum_{i,p} (\lambda^i_p+e^i_p)^2+c.c. -2k) 
\prod_{p} \exp(-\frac{N Z^{-1}(2 \pi)^2}{g_W^2}(\sum_{i} (\lambda^i_p+e^i_p)^2+c.c.)) \nonumber \\
\tilde a^{-2Nk}\prod_{i > j} ((\lambda_p ^{i} - \lambda_p ^{j} +e_p ^{i } -e_p ^{j })
\times c.c.)
\eea
and
\bea
\exp(-\Gamma^{k,0}_{q})= \sum_m \delta(\sum_{i,p}(\lambda^i_p+m^i_p)-k) 
\delta(\sum_{i,p}(\bar \lambda^i_p+\bar m^i_p)+k) \nonumber \\
 \delta(\sum_{i,p} (\lambda^i_p+m^i_p)^2+c.c.-2k) 
\prod_{p} \exp(-\frac{N (2 \pi)^2 \tilde a^2}{a^2 g_W^2}(\sum_{i}  (\lambda^i_p+ m_p ^{i })^2+c.c.))
\nonumber \\
\prod_{i >j} ((\lambda_p ^{i} - \lambda_p ^{j} +m_p ^{i } -m_p ^{j })\times c.c.)
\eea
The complete expression, as a sum over $k$ sectors, is:
\bea
\sum_k \exp(-\Gamma^{k,\infty}_{q}) \exp(-\Gamma^{k,0}_{q})
\eea
If we set in canonical form the quadratic part of the action
in the two holograms, we get a different rescaling for different
fields. For the hologram at infinity we get:
\bea
(g_c Z^{\frac{1}{2}})^{2Nk}
\eea
The exponent, $2Nk$, results from the number of zero modes, that equals the complex dimension of the adjoint
orbit for vortices in the fundamental representation, $Nk$, plus the conjugate contribution, $Nk$,
that in turn equals the number of distinct factors $\lambda_p ^{i} - \lambda_p ^{j} +e_p ^{i } -e_p ^{j }$
in the Vandermonde determinant times complex conjugate contribution. Had we employed the resolution of
identity in
Eq.(20), we would have obtained twice as many factors, but at the same time twice as many zero modes,
because of the different holomorphic anti-holomorphic pairing. However, in this case, the normalisation
of the classical action would have been different, in order to get the correct one- and two-loop beta function.
The canonical rescaling of the action in the hologram at infinity
involves, in addition to the usual rescaling by a factor of $g$,
also the multiplicative renormalization, $Z$,
of the square of the $ASD$ curvature, as opposed to the $SUSY$ case.
For the hologram at zero, by the same counting, we get instead:
\bea
(g_c)^{2Nk}
\eea
since the multiplicative renormalization, $Z$, is finite on the hologram at zero and it can be set 
equal to one by a wise choice of the subtraction point.
Putting together the two contributions, we get:
\bea
\frac{1}{2 g^2_W} 
 = \frac{1}{2 g^2_c}+ \beta_J log g_c + \frac{\beta_J}{4} logZ
\eea
in the continuum limit, defined as $\tilde a \rightarrow 0, a=const$. 
From Eq.(84) it follows the formula of $NSVZ$ type \cite{Shiff} for the canonical coupling
mentioned in the introduction:
\bea
\frac{\partial g_c}{\partial log \Lambda}=\frac{-\beta_0 g_c^3+
\frac {\beta_J}{4} g_c^3 \frac{\partial log Z}{\partial log \Lambda} }{1- \beta_J g_c^2 }
\eea
We expect the existence of a (scheme dependent) infrared fixed point
due to cancellations of terms of different order in $g_c$ in the numerator, since,
contrary to perturbation theory,
the order of $g^5$ contribution in the numerator has the sign opposite to the one of order of $g^3$.

\section{Conclusions}

We have computed exactly, up to finite terms, the local part of the effective action 
for the eigenvalues of the curvature of a twisted local system, that solves the 
holographic loop equation, for a certain self-avoiding Wilson loop in the adjoint representation,
in the large-$N$ limit of a twisted Eguchi-Kawai reduced version of $QCD_4$.
The construction of the effective action employs a hologram made by two charts, with the loop on their boundary,
containing two lattices of points, on which the curvature of the twisted local system
is localised. The hologram is a conformal image of the two regions of the plane delimited
by the loop, with attached infinitesimal strips ending into cusps at infinity, that are the
conformal images of the points of the lattice in each chart.
On the hologram the theory becomes classical in the large-$N$ limit, in the sense that the
quantum term vanishes in the loop equation, because of the zig-zag symmetry in a neighbourhood of the cusps,
and hence the holographic loop equation is implied by the critical equation of the effective action.
We have extracted from it the exact
beta function for the Wilsonean and the canonical coupling.
In a certain scheme, the Wilsonean coupling is only one loop,
with the correct perturbative first coefficient of the beta function.
In the same scheme, the canonical coupling has a $NSVZ$ form, 
with the correct first two universal perturbative coefficients of the beta function.
In addition, the exact canonical beta function may have a scheme dependent fixed point in the infrared,
due to the opposite signs in the numerator in the fraction of $NSVZ$ type.
We leave for the future the study of the inter-quark potential via the effective action.

\end{document}